\documentstyle[aps,prd,epsf,floats]{revtex} 
\draft

\begin{document}
\twocolumn[\hsize\textwidth\columnwidth\hsize\csname
@twocolumnfalse\endcsname 

\begin{flushright}
UCLA/00/TEP/2
\end{flushright}

\preprint{UCLA/00/TEP/2} 

\title{Baryon number non-conservation and phase transitions at preheating
}

\author{John M. Cornwall$^1$ and Alexander Kusenko$^{1,2}$}
\address{$^1$Department of Physics and Astronomy, UCLA, Los Angeles, CA
90095-1547, USA \\ $^2$RIKEN BNL Research Center, Brookhaven National
Laboratory, Upton, NY 11973, USA}

\date{\today}

\maketitle
             
\begin{abstract}

Certain inflation models undergo pre-heating, in which inflaton
oscillations can drive parametric resonance instabilities.  We discuss
several phenomena stemming from such instabilities, especially in
weak-scale models; generically, these involve energizing a resonant system
so that it can evade tunneling by crossing barriers classically.  One
possibility is a spontaneous change of phase from a lower-energy vacuum
state to one of higher energy, as exemplified by an asymmetric double-well
potential with different masses in each well. If the lower well is in
resonance with oscillations of the potential, a system can be driven
resonantly to the upper well and stay there (except for tunneling) if the
upper well is not resonant. Another example occurs in hybrid inflation
models where the Higgs field is resonant; the Higgs oscillations can be
transferred to electroweak (EW) gauge potentials, leading to rapid
transitions over sphaleron barriers and consequent B+L violation.  Given an
appropriate CP-violating seed, we find that preheating can drive a
time-varying condensate of Chern-Simons number over large spatial scales;
this condensate evolves by oscillation as well as decay into modes with
shorter spatial gradients, eventually ending up as a condensate of
sphalerons.  We study these examples numerically and to some extent
analytically.  The emphasis in the present paper is on the generic
mechanisms, and not on specific preheating models; these will be discussed
in a later paper.

\end{abstract}


\vskip2.0pc]


\renewcommand{\thefootnote}{\arabic{footnote}}
\setcounter{footnote}{0}

\section{Introduction}
\label{sec-1}

There are well-known reasons to believe that inflation took place and was
followed by reheating to some temperature $T_{_R}$.  Before a thermal
equilibrium was reached, the coherent oscillations of the inflaton could
create the environment in which a resonant non-thermal production of
particles could rapidly transfer energy from the inflaton to the other
fields.  This stage, known as preheating~\cite{preheating}, has been a
subject of intense studies.  In particular, it was argued that both
non-thermal phase transitions~\cite{pt} and the generation of baryon
asymmetry~\cite{pbg,kt,ggks} could occur during preheating.

We will describe two new field-theoretical phenomena that can be caused by
coherent oscillations of the inflaton. One is a new example of a phase
transition driven by the coherent oscillations of the inflaton.  This
transition has an unusual feature that it can start in a lower-energy
ground state and end in a higher-energy metastable vacuum. We discuss this
in Section II. 

In Section III we describe resonant generation of a fermion density through
anomalous gauge interactions that can be the basis for baryogenesis.  In
contrast with the earlier work, where the analyses were
based on analogies with thermal sphalerons~\cite{ggks,gg} or topological
defects~\cite{kt}, we construct an explicit solution that can be though of
as a condensate of sphalerons.  We show that the evolution of this solution
can lead to a resonant growth of Chern-Simons number density.

\section{Phase transitions at preheating}

The properties of the physical vacuum and the particle content of the
universe are determined by physical processes that took place in a hot
primordial plasma.  Theories of particle interactions beyond the Standard
Model allow for different types of physical vacua.  For example, an SU(5)
Grand Unified Theory (GUT) allows three possibilities for the ground state,
in which the gauge group that remain unbroken is, respectively, SU(5),
SU(4)$\times$U(1), or SU(3)$\times$SU(2)$\times$U(1).  If low-energy
supersymmetry is assumed (to assure the gauge coupling unification and to
stabilize the hierarchy of scales), these three ground states are
degenerate in energy up to small supersymmetry breaking terms
$\sim$~TeV.  Therefore, any of these potential minima could equally well
be the present physical vacuum.  The evolution of the universe shortly
after the Big Bang must have chosen SU(3)$\times$SU(2)$\times$U(1) vacuum
over the others.  The phenomenon we will discuss can provide a new solution
to the old puzzle related to breaking of a SUSY GUT gauge group.  The same
process can have important consequences in other models with several
competing (metastable) vacua, for example, in the minimal supersymmetric
extention of the Standard Model (MSSM).

Let us consider an inflaton $\Phi$ interacting with a ``Higgs field''
$\chi$ through a coupling of the form $\lambda \Phi^2 \chi^{\dag} \chi$ or
$\mu \Phi \chi^{\dag} \chi$, or both.  Let us assume that the effective
potential $V(\chi,\Phi)$ has two non-degenerate minima, for example at
$\langle \chi \rangle =\pm v$, $\langle \Phi \rangle =v_{_I}$, and that the
mass of the $\chi$ particle is not the same in both minima, that is
$\partial^2 V(v,v_{_I})/\partial \chi^2 \neq \partial^2
V(-v,v_{_I})/\partial \chi^2$.

\begin{figure}[t]
\centering
\hspace*{-5.5mm}
\leavevmode\epsfysize=6.5cm \epsfbox{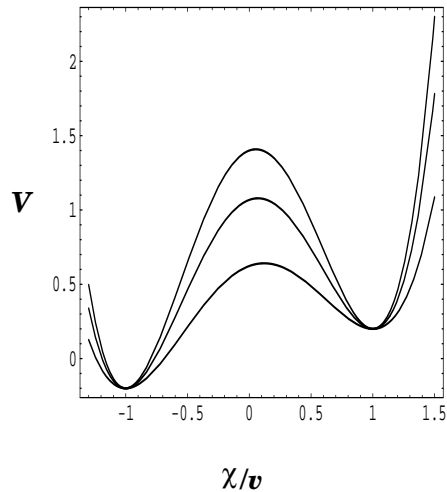}\\[3mm]
\caption[fig1]{\label{fig1} The time-dependent potential {$V(\chi,\Phi(t))=
(\chi^2-v^2)^2 [1+0.4 \cos 5.6 v t]+0.1 v \chi (3 v^2-\chi^2) $} that has
two non-degenerate minima and a time-dependent barrier height.  The masses of
the {$\chi$} particles are also time-dependent and are different in the two
minima.  
}
\end{figure}

\begin{figure}[t]
\centering
\hspace*{-5.5mm}
\leavevmode\epsfysize=6.5cm \epsfbox{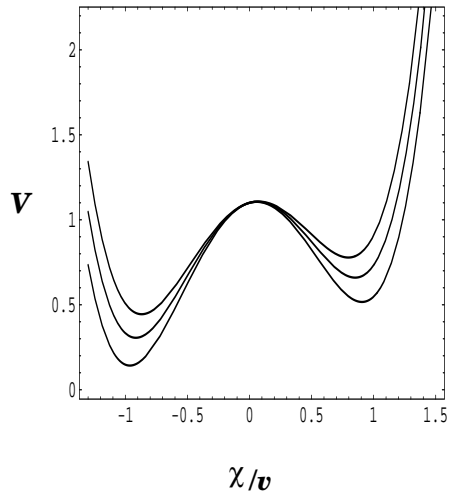}\\[3mm]
\caption[fig2]{\label{fig2} 
Another time-dependent potential.  The heights of the two vacua oscillate. 
}
\end{figure}

At the end of inflation, the system can occupy the lowest-energy state
with $\langle \chi \rangle =- v$.  During preheating, the inflaton
oscillates around its VEV, $\Phi(t) = v_{_I} + \Phi_0 \cos \omega t$.  In
general, this induces a time-dependent mass for the Higgs field $\chi$
through the couplings $\mu$ and $\lambda$.  The equation of motion for
the homogenous (zero-momentum) mode of the field $\chi$ is 
\begin{equation}
\ddot{\chi} + 3 H \dot{\chi} + \frac{\partial}{\partial \chi} V(\chi, v_{_I}
+ \Phi_0 \cos \omega t) =0, 
\label{eqn_motion}
\end{equation}
where $H$ is the Hubble constant.\footnote{In weak-scale preheating the
Hubble constant is negligiblly small.  For GUT-scale preheating it is not,
and it could play an important role in helping to scan resonant bands.} In
Fig.~\ref{fig1} and Fig.~\ref{fig2} we show two examples of time-dependent
effective potentials.

The potential $V(\chi,\Phi(t))= (\chi^2-v^2)^2 [1+0.4 \cos 5.6 v t]+0.1 v
\chi (3 v^2 -\chi^2) $ depicted in Fig.~\ref{fig1} has two classical solutions,
$\chi=-v$ and $\chi=v$.  Naively one could expect that the lowest-energy
solution $\chi=-v$ corresponds to the vacuum state.  This is not
necessarily the case, however.  Since the mass of the $\chi$ field is
time-dependent, the solution $\chi (t) =-v$ may be unstable with respect to
small perturbations.  At the same time, the other solution, $\chi (t) =+v$
may be stable. If this is the case, the classical system is attracted to
the trajectory $\chi(t) =+v$.

\begin{figure}[t]
\centering
\hspace*{-5.5mm}
\leavevmode\epsfysize=6.5cm \epsfbox{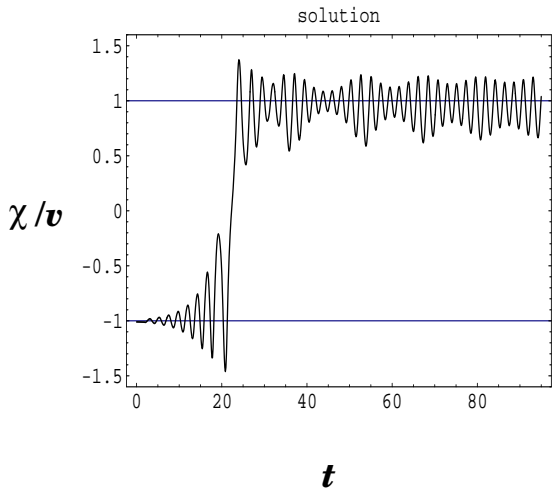}\\[3mm]
\caption[fig3]{\label{fig3} Classical solution of the equation of motion in
the potential of Fig.~\ref{fig1}.  The evolution begins near the unstable
classical trajectory {$ \chi(t)=-v $} and is driven towards a stable
classical solution {$ \chi(t)=+v $}. In quantum theory, if the tunneling
rate between the two vacua is small, a phase transition to a metastable
vacuum takes place.  }
\end{figure}

In the vicinity of the global minimum, for $|(\chi + v)/v| \ll 1$, the
equation of motion (\ref{eqn_motion}) is a Mathieu equation that has
rapidly growing solutions for some values of $\omega$, $\Phi_0$, and $
m^{(-)} \equiv \partial^2 V(-v,v_{_I})/\partial \chi^2 $.  The inflaton
frequency changes with time and can enter in resonance, at which point
$(|\chi (t)| - v)$ begins to grow exponentially.  This kind of solution of
equation (1), with $H$=0 and the potential of Fig.~\ref{fig1}, is shown in
Fig.~\ref{fig3}.  At some point it crosses the barrier and begins
oscillations around a different potential minimum, $\langle \chi \rangle =
+ v$.  However, the mass of the $\chi$ particle near $\langle \chi \rangle
= + v$ is $m^{(+)}$, different from $m^{(-)}$.  Therefore, the system may
go out of resonance after crossing the barrier.  There are no growing
solutions in the vicinity of the seconds minimum, and the oscillations die
out with $\langle \chi \rangle = + v$.

If the tunneling rate between $\langle \chi \rangle = + v$ and $\langle
\chi \rangle = - v$ is negligible, the classical evolution shown in
Fig.~\ref{fig3} describes a phase transition into a metastable false vacuum. 

This example shows that the ground state at the end of inflation does not
necesarily correspond to the global minimum of the potential.  Instead,
during the preheating, a false vacuum can be populated if the true vacuum
entered in resonace while the false vacuum did not.  

Both Grand Unified Theories and supersymmetric extentions of the Standard
Model predict the existence of local minima in the effective potential.
The tunneling rate between these minima can be extremely low and their
lifetimes can easily exceed the present age of the universe.  For example,
the effective potential of the MSSM can have a broken color SU(3) in its
global minimum, while the standard, color and charge conserving vacuum is
metastable.  For natural and experimentally allowed values of the MSSM
parameters, the lifetime of this false vacuum can be much greater than
$10^{10}$ years~\cite{kls}.  If the reheat temperature after inflation was
not much higher than the electroweak scale, this metastable minimum could
be populated in the way we have described. 

Breaking a SUSY GUT gauge group and choosing between the nearly degenerate
minima is problematic in non-inflationary cosmology~\cite{nilles}.  Let us
consider a SUSY SU(5) GUT for example.  The minima with unbroken SU(5),
SU(4)$\times$U(1), and SU(3)$\times$SU(2)$\times$U(1) groups are nearly
degenerate, split only by supersymmetry breaking terms of the order of a
TeV.  Why did the universe end up in the vacuum with the lowest symmetry? 

Finite temperature corrections (if relevant, which may not be the case for
preheating) make the SU(5) minimum lowest in energy 
because it has a higher number of degres of freedom.  The subsequent
thermal evolution of the potential makes tunneling into a Standard Model
vacuum impossible~\cite{nilles} even if it becomes the global minimum at
temperatures below 1~TeV.  Supergravity splits the three minima by a
negligible amount and in such a way that cosmological constant can by
fine-tuned to zero only in the minimum with the higher energy while the
other two minima have negative energy density~\cite{weinberg}.  Some of the
proposed solutions~\cite{nilles} rely on assumptions about a strong gauge
dynamics that seem somewhat implausible.

If, however, inflation took place, the SUSY GUT vacuum could be chosen in a
phase transition of the kind we described.  This appears to resolve a
long-standing problem concerning the breaking of the SUSY GUT gauge group. 

\section{B+L violation} 

As discussed in the Introduction, preheating oscillations of the Higgs VEV
can lead to two effects of interest for B+L violation.  The first \cite{gg}
is that the sphaleron barrier itself oscillates, leading in principle to
exponentially-sensitive oscillations of the sphaleron rate.  The second,
which we take up here, is that Higgs oscillations can resonantly drive
classical transitions over the barrier.

Given an appropriate CP-violating seed, there are three stages to this
classical resonant driving.  In the first stage, the seed (which can be a
source term or initial conditions on the EW gauge potentials) drives
large-scale generation of Chern-Simons (CS) number (topological charge)
over spatial scales so large that spatial variation can be ignored and only
temporal variation saved in the classical equations of motion.  In the
second stage, gradients on shorter scales emerge, as a result of unstable
growth of spatially-dependent perturbations.  The seeds for these spatial
modes might emerge from spinodal decomposition during inflation
\cite{ch,tt}.  As expected on general grounds from earlier preheating
studies, the fastest-growing modes are those with large spatial scales. The
third stage involves the generation of sphalerons, with spatial scales at
the standard W-boson mass $M_W$.

In all stages, we will ignore various back-reaction effects; the expansion
of the universe (in any case, neglible for weak-scale inflation); and
damping produced by perturbative decays (one order of $\alpha_W$ higher
than terms we keep).

We discuss the first stage, which has important non-linear effects stemming
from gauge-potential self-coupulings, both analytically and numerically.  A
particular {\it ansatz} is used for the gauge potential, having only a
time dependence.  (This {\it ansatz} has been used some time ago
\cite{co90} in a rather different scenario.)  The analysis is in the same
spirit as the conventional approach to low-order resonances in the Mathieu
equation (see, {\it e.g.}, Ref. \cite{ll}).  But the lowest-order
resonant-mode equations, two first-order differential equations, have a
cubic non-linearity.  Surprisingly, these coupled non-linear equations can be
solved exactly in terms of elliptic functions.  The non-linear terms not
only provide a quartic potential opposing the growth of CS number but, as
the CS number grows, the non-linear term also grows and drives the system
off resonance.  In effect, the cubic non-linearity causes the W-boson mass
to increase.  Interestingly, this increase can be offset by a secular
increase of the frequency of Higgs oscillations, allowing resonance to be
maintained for long periods of time with consequent large growth of CS
number.

In the second stage we include linear perturbations to the
spatially-homogeneous equations of the first stage; these perturbations are
considered to lowest order in spatial gradients, as characterized by a
spatial momentum $k$.  It is not possible to do a conventional
dispersion-relation analysis of these equations, which have time-dependent
coefficients as determined by the temporal growth of the first-stage gauge
potentials.  We perform a numerical analysis of the three coupled linear
differential equations which result.

The third stage, in which gradients evolve to spatial scales $\sim
M_W^{-1}$ appropriate for sphalerons, is the hardest to analyze, since an
adequate treatment involves the solution of coupled partial differential
equations with time-dependent coefficients.  So we restrict ourselves to a
crude, simple first step, reducing these partial differential equations by
a non-linear ordinary differential equation for an approximate
sphaleron-like mode.  The relevant gauge-potential {\it ansatz}, first
introduced by Bitar and Chang \cite{bc}, was later used \cite{co89} to
analyze sphalerons above the EW phase transition, and was shown to have an
effective barrier potential for the sphaleron which was numerically very
close to that of a simple pendulum.  We introduce an oscillating Higgs
field, which causes this pendulum to be parametrically-driven.  The {\it
ansatz} is too simple to be used for anything more than estimating the rate
of change of topological charge as the pendulum goes over its barrier once;
we do this numerically.  In principle, more complicated forms, representing
multiple sphalerons, could be used, such as the ADHM construction or those
of 't~Hooft or of Jackiw, Nohl, and Rebbi \cite{jnr} multi-instanton form,
suitably modified for Minkowski-space dynamics, but these have not yielded
any insights for us. 

At all stages, the energy density associated with generation of CS number
is of order $4\pi m^4/g^2$, as would be appropriate for a gas of sphalerons
with density $\sim m^3$.

\subsection{First stage:  homogeneous CS parametric resonance} 

In what follows we always consider the Higgs field to have a given VEV, as
determined by preheating effects.  Introduce the conventional
anti-hermitean gauge potential, with coupling $g$ included, by:
\begin{equation}   
gA_{\mu}=(\frac{\tau_a}{2i})A_{\mu}^a.
\end{equation}   
Our spatially-homogeneous {\it ansatz} is:
\begin{equation}   
gA_0=0;\;gA_i=(\frac{\tau_i}{2i})\phi(t)
\end{equation}
in which the group index is tied to the spatial index.  By the conventional
rules of charge conjugation and parity for the gauge potential, $\phi$ is C
even, P odd, CP odd.

It is important to note that this {\it ansatz} does not correspond to a
non-vanishing VEV for an EW field.  Gauge invariance alone is enough to
ensure that there can be no expectation value coupling the space-time
indices to group indices.

One readily calculates the EW electric and magnetic fields:
\begin{equation}   
gE_i\equiv G_{0i}=(\frac{\tau_i}{2i})\dot{\phi}(t);\;
gB_i\equiv \frac{1}{2}\epsilon_{ijk}G_{jk}=(\frac{\tau_i}{2i})\phi^2.
\end{equation}
Then one calculates the density $W$ of Chern-Simons number as:
\begin{equation}     
W=(\frac{1}{8\pi^2})\phi^3.
\end{equation}
It is straightforward to check that $\dot{W}$ is the topological charge
density $Q$, related to B+L violation through the anomaly equation.

With the assumption of a given Higgs VEV, the equations of motion for the
gauge potential are:
\begin{equation}     
[D^{\mu},G_{\mu\nu}]+M_W^2(t)(A_{\nu}+(\partial_{\nu}U)U^{-1})=0.
\end{equation}
Here the unitary matrix $U$ represents the Goldstone (phase) part of the
Higgs field.  The mass term will be assumed to have the form:
\begin{equation}     
M_W^2(t)=m^2(1+\epsilon \cos (\omega t))
\end{equation}
where $m$ is the value of $M_W$ with no oscillations.  Later we will have
occasion to consider a time-dependent frequency $\omega$, but for now
think of it as a constant.

There must be some sort of CP-violating seed to produce non-zero solutions
of the equations of motion; these might stem from (long-scale) spatial
gradients in the matrix $U$, which acts as a source in equation (6), or
from initial values of $\phi$.  Because the equations are unstable, there
is little practical difference, and we choose to drop the $U$ terms in the
equations of motion, and then providing a seed through initial values.
Then there is a single equation for $\phi$:
\begin{equation}     
\ddot{\phi}+2\phi^3+(1+\epsilon \cos r t)\phi =0
\end{equation}  
We have non-dimensionalized the equations of motion by measuring $\phi$ in
units of $m$ and time $t$ in units of $m^{-1}$.  The parameter $\rho$ has
the value $\omega /m$.

Without the cubic non-linearity, this would be a standard Mathieu equation.
In the Appendix we analyze the coupled non-linear mode equations which
arise for the lowest resonance ($r=2$), and find that they can be solved
exactly in terms of elliptic integrals.  The qualitative features of this
analysis are easy to anticipate: Equation (8) describes the motion of a
particle in a quartic potential. The oscillating term drives the particle
up the wall, but eventually the particle gets out of resonance and falls
back.  This process can repeat quasi-periodically.

\begin{figure}[t]
\centering
\hspace*{-5.5mm}
\leavevmode\epsfysize=6.5cm \epsfbox{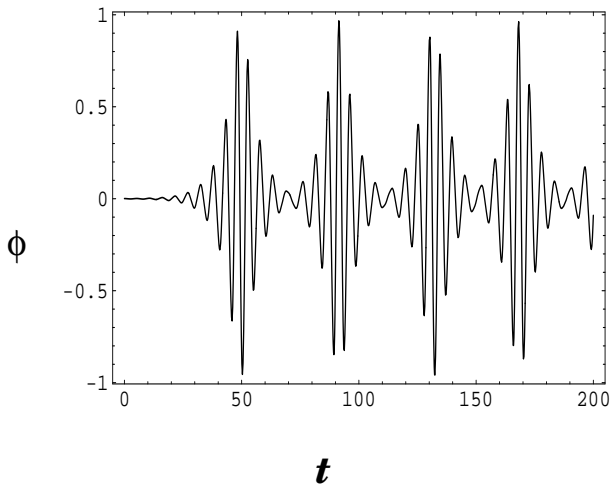}\\[3mm]
\caption[fig4]{\label{fig4} Time dependence of $\phi$ for initial values as
in equation (9) plus $r=2.3,\epsilon =0.9$.  
}
\end{figure}

We now turn to numerical analysis.  Only a couple of examples will be
reported, without attempting to choose parameters to correspond to
realistic preheating scenarios.  Parameters are chosen to illustrate
specific effects; other parameter sets may show no interesting behavior at
all.  The runs reported here have initial values 
\begin{equation}    
\phi(0)=0.001;\;\dot{\phi}(0)=0,
\end{equation}
and large values of $\epsilon$, in the range 0.5-0.9.  
Because the equations are both non-linear and unstable, the final results
are largely independent of the initial conditions as long as they are
non-zero.  As the initial values are 
reduced, the time of onset of instability is sometimes lengthened.
Generally, there are two regimes (for constant frequency $\omega$): The
resonant regime, in which $\phi$ 
grows to O(1), and the non-resonant regime where $\phi$ stays small.  We
will only show the near-resonant cases in the figures.  There is another
regime in which $\omega$ grows secularly with time, and which leads to
larger values of $\phi$.  

Fig.~\ref{fig4} is a typical example of the behavior when $\omega$ or $r$
is constant and fairly near resonance (in this case, $r$=2.3).  One sees
that the envelop of $|\phi|$ grows to order unity, but periodically passes
through zero and repeats.  This is because $\epsilon$ is near unity, and so
system frequencies vary quite a bit, from $1+\epsilon$ to $1-\epsilon$. 

\begin{figure}[t]
\centering
\hspace*{-5.5mm}
\leavevmode\epsfysize=6.5cm \epsfbox{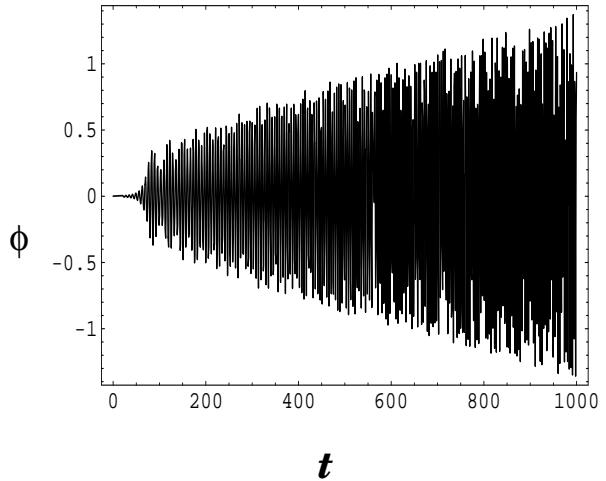}\\[3mm]
\caption[fig5]{\label{fig5} 
Behavior of $\phi$ with the initial conditions of equation (9), with
$b=0.52$, and a secularly-growing frequency $r(t)=1.7+0.001t$.  
}
\end{figure}

Fig.~\ref{fig5} shows the behavior when the frequency grows secularly.  The
onset of rapid growth is delayed because the system is originally fairly
far from resonance, but then the envelop of $|\phi|$ grows essentially
linearly, coupled to the frequency change.  The system is able to stay in
resonance as $\phi$ grows linearly, because the effective mass $M$ of the
$\phi$ field (see the Appendix) is $M^2\simeq m^2+3\langle \phi^2\rangle$,
and the effective ratio $r=\omega /M$ stays roughly constant if $M$ grows
at the same rate as $\omega$. 

\begin{figure}[t]
\centering
\hspace*{-5.5mm}
\leavevmode\epsfysize=6.5cm \epsfbox{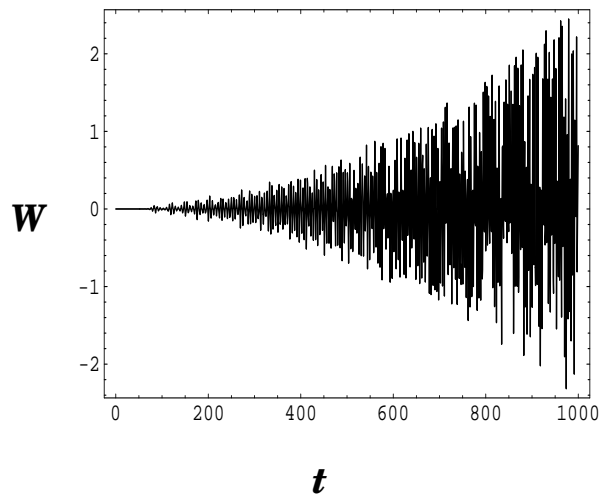}\\[3mm]
\caption[fig6]{\label{fig6}
The Chern-Simons density (see equation (5)) for the parameters of
Fig.~\ref{fig5}.} 
\end{figure}

Fig.~\ref{fig6} shows the CS density $\phi^3/8\pi^2$ corresponding to the
parameters of Fig.~\ref{fig5}.  The CS density grows roughly as $t^3$, with
$\phi$ growing linearly in time as does $\omega$.    

With dimensionalized values of $|\phi| \simeq m$, the CS number density is
of order 0.01 $m^3$, corresponding to a large B+L density.  Whether any of
this CS density survives preheating to the reheating phase depends on
whether there is a ``graceful exit" to preheating generation of CS number,
and this depends on factors not considered in this paper, such as back
reaction, growth of finite-momentum modes, and linear damping by decay of
the W-boson condensate.  Additionally, there may be many domains large
compared to $m^{-1}$ but small compared to the Hubble size in which the
values of $\phi$ are uncorrelated.  This will reduce the effective global
CS density by a factor of $N^{1/2}$, where $N$ is the number of such
domains.  The ultimate fate of the processes considered here will be taken
up in a future work, in which specific weak-scale preheating scenarios will
be taken up.

\subsection{Second stage:  evolution of spatially-varying modes}

Ultimately, there will be some CP-violating seeds with finite spatial
gradients.  Assuming that these seeds are smaller than those for $\phi$ (as
is reasonable following inflation), these seeds will be driven by the time
variation of $\phi$ as well as of the Higgs VEV.  We will be concerned here
only with the linearized equations for the spatially-varying modes, which
we characterize in momentum space. As is usual in preheating phenomena, the
modes with the longest spatial scales (smallest $k$) grow fastest.

The total vector potential is written as $A_{\mu}+a_{\mu}$, with $A_{\mu}$
taken from equation (3).  The most general vector potential $ga_{\mu}$
depending on a single vector $\vec{k}$ has time component
\begin{equation}    
ga_0=(\frac{i\vec{\tau}\cdot \hat{k}}{2i})\alpha_0,
\end{equation}
and space components
\begin{equation}     
ga_j=\frac{1}{2i}[(\tau_j-\hat{k}_j\vec{\tau}\cdot
\hat{k})\beta_1+i\epsilon_{jab}\tau_a\hat{k}_b\beta_2+\hat{k}_j\vec{\tau}\cdot
\hat{k}\beta_3].
\end{equation}
In equations (10,11) the hat indicates a unit vector, and
$\alpha_0,\beta_i$ are real functions of $k^2$ and $t$.  As before, we
non-dimensionalize by dividing these functions by $m$, replacing $t$ by
$mt$, and $k$ by $k/m$.  Presumably the Fourier transforms in (10,11)
vanish at an appropriate rate as $k\rightarrow 0$ so as to change $\hat{k}$
into $\vec{k}$, although this will not matter in what follows.

It is straightforward if lengthy to write out the linearized version of
equation (6) (without the $U$ terms):
\begin{eqnarray}   
\alpha_0 & = &
\frac{1}{Q}[2(\dot{\phi}\beta_2-\dot{\beta}_2\phi)-k\dot{\beta}_3],\\
\nonumber Q & = & k^2+2\phi^2+1+\epsilon \cos rt;
\end{eqnarray}
\begin{equation}     
\ddot{\beta}_1+Q\beta_1-2k\phi \beta_2 +2(\beta_1+\beta_3)\phi^2=0;
\end{equation}
\begin{equation}    
\ddot{\beta}_2+Q\beta_2-2k\phi \beta_1 +\phi (\dot{\alpha}_0-k\beta_3
)+2\dot{\phi}\alpha_0=0;
\end{equation}
\begin{equation}   
\ddot{\beta}_3+Q\beta_3+k(\dot{\alpha}_0-k\beta_3-2\phi
\beta_2)+4\beta_1\phi^2=0.
\end{equation}

\begin{figure}[t]
\centering
\hspace*{-5.5mm}
\leavevmode\epsfysize=19.5cm \epsfbox{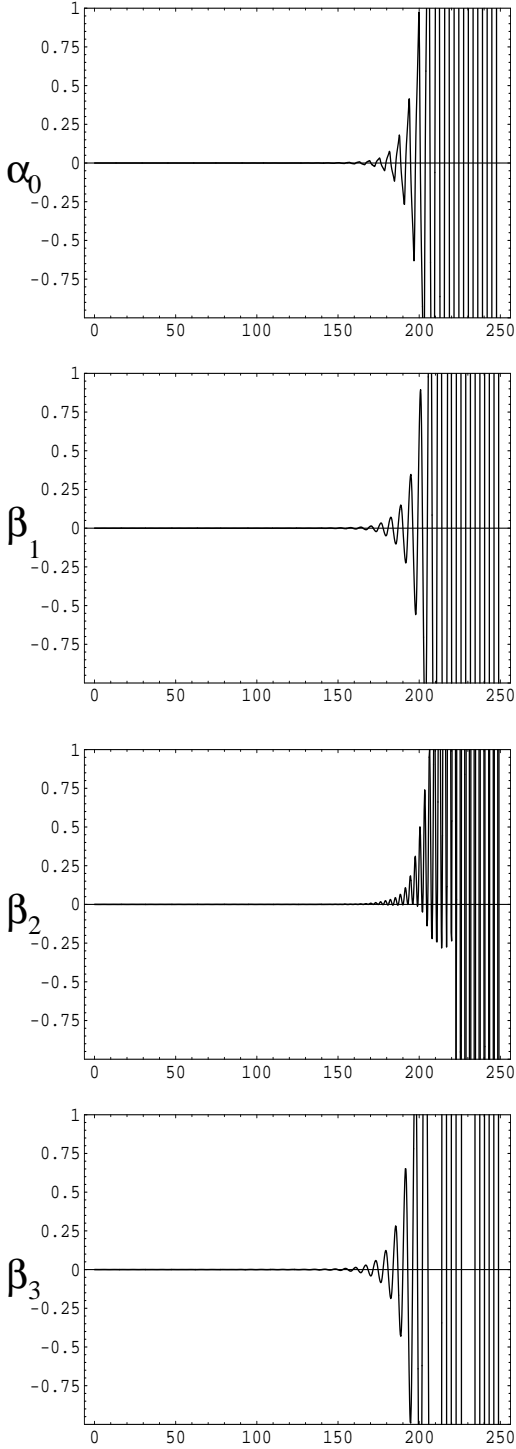}\\[3mm]
\caption[fig7]{\label{fig7} 
The evolution of the linear modes for the
parameters of Fig.~\ref{fig5}.  Note the truncation on the vertical axis;
at $t\simeq$ 200} all the amplitudes are larger than one in magnitude.
\end{figure}

Even though these are linear equations for the modes $\alpha_0,\beta_j$
they are impossible to solve analytically, because $\phi$ is not an
analytically-known function.  We have solved them numerically, with various
interesting results.  Perhaps the most interesting is that these mode
functions remain small and well-behaved for a long time, and then when
$\phi$ is large enough (of order unity) they show violently unstable
behavior.  This is especially so for the case when the frequency $\omega$
is growing with time, as for Fig.~\ref{fig5}.  This is illustrated in
Fig.~\ref{fig7}, showing the evolution with time of the linear modes for
the parameters of Fig.~\ref{fig5}.  The mode functions were begun with
initial values which are 0.1 times those of $\phi$ (see equation (9)).  Of
course, any other initial values can be gotten by scaling, since the
equations are linear.  The point is that when, for a given set of initial
values of $\alpha_0,\beta_j$, these functions rise to be of O(1), the whole
problem becomes non-linear and presumably enters something like the third
stage discussed below.  Note in Fig.~\ref{fig7} that the threshhold for
non-linearity, with the given initial conditions, occurs at a
(dimensionless) time of 200, which gives $\phi$ enough time to get big
enough to be interesting (see Fig.~\ref{fig5}).

\subsection{Third stage:  sphalerons}

Eventually, momentum modes with $k\simeq 1$ will become prominent, and the
condensate of CS number becomes a condensate of sphalerons.  It is much
more difficult to describe this stage, and we will only take a simple first
step. This step consists of a drastic simplification of the kinematics of a
sphaleron coupled to a time-dependent Higgs field, reducing the dynamics to
a single function $\lambda (t)$ as in Refs. \cite{bc,co89}.  Write the most
general spherically-symmetric gauge potential and Higgs phase matrix $U$ in
the form:\
\begin{equation}   
U=\exp [\frac{i\gamma }{2} \hat{r}\cdot
\vec{\tau}],\;gA_0=\frac{1}{2i}\hat{r}\cdot \vec{\tau}H_2;
\end{equation}
\begin{equation}    
gA_i= \frac{1}{2ir}
[\epsilon_{iak}\tau_a\hat{r}_k(\phi_1-1)-(\tau_i-\hat{r}_i\hat{r}\cdot
\vec{\tau})\phi_2+\vec{r}_i\hat{r}\cdot \vec{\tau}H_1.
\end{equation}
The functions $H_i,\phi_j$ depend only on $r,t$.  The asymptotic values of
the angle $\gamma$ are zero at $r=0$ and $\pi$ at $r=\infty$.  We
parametrize these functions as:
\begin{eqnarray}  
H_1 & = & \frac{2\lambda}{\lambda^2+r^2+a^2}; \\ \nonumber H_2 & = &
 -\frac{2r\dot{\lambda}}{\lambda^2+r^2+a^2};\\ \nonumber \phi_1 & = &
 1-\frac{2r^2}{\lambda^2+r^2+a^2}; \\ \nonumber \phi_2 & = &
 -\frac{2r\lambda }{\lambda^2+r^2+a^2}.
\end{eqnarray}
For details on the parametrization of $\gamma$ see \cite{co89}. For the
present purpose one can just think of $\gamma$ as always equal to $\pi$.
In this parametrization the constant $a$ is a size parameter (like that of
an instanton) and $\lambda$, the sole dynamic degree of freedom, depends on
$t$.  Generally, $\lambda$ is an odd function of $t$, vanishing along with
its first derivative at $t=0$.

The electric and magnetic fields are:
\begin{eqnarray}
gE_j & = & (\frac{\tau_j}{2i})\frac{4a^2\dot{\lambda}}{\lambda^2+r^2+a^2};
\\ \nonumber gB_j & = & (\frac{\tau_j}{2i})\frac{4a^2}{\lambda^2+r^2+a^2}.
\end{eqnarray}
Note that these have the same space and internal symmetry index dependence
as does the $\phi$ {\it ansatz} of equation (3).  It is therefore natural
to suppose that the $\phi$ fields will transform (through the growth of
spatial modes) into a condensate of sphalerons.  Of course, in this
condensate each sphaleron will be a translate in space and in time of the
sphaleron exhibited here, which is centered at the space-time origin.

With boundary conditions
\begin{equation}    
\lambda (t=-\infty ) =-\infty ;\;\lambda (t=+\infty )=+\infty 
\end{equation}
one readily verifies that, no matter what the dynamics of $\lambda$ as long
as it is single-valued, the (Minkowskian) topological charge
\begin{equation}    
Q=-\frac{g^2}{4\pi^2}\int d^4x Tr\vec{E}\cdot \vec{B}
\end{equation}
has the value 1. Indeed, if we replace $\lambda$ by $t$ we get exactly the
usual Euclidean one-instanton expression, which however is now being
interpreted as a Minkowskian construct.

The size coordinate $a$ is not arbitrary, as it is for instantons in gauge
theories with no Higgs field.  As shown in \cite{co89}, if one goes to
$t=0$ and sets $\lambda ,\dot{\lambda}=0$ there, the resulting $ansatz$ in
equations (16,17) is an excellent trial wave function for the sphaleron.
Minimizing the Hamiltonian (for time-independent Higgs VEV) yields $a=\sqrt
3/2M_W)$ and a sphaleron mass $M_s$ only a fraction of a percent higher
than the true value, determined numerically, of
\begin{equation}       
M_s=5.41(\frac{4\pi M_W}{g^2}).
\end{equation}
When the mass $M_W$ depends on time, as in equation (7), we will continue
to use the above value for $a$.  It then happens that the parameters of the
Hamiltonian depend on time (see \cite{co89} for the Hamiltonian as a
function of $a,\lambda ,\dot{\lambda}$).

As is further shown in \cite{co89}, one can trade the function $\lambda$
for a topological charge $Q(t)$ defined by demanding that the kinetic
energy term in the Hamiltonian is of the form $(1/2)I\dot{Q}^2$ with $I$
independent of $Q$.  The normalization
\begin{equation}     
\lambda = -\infty : \; Q=0;\;\lambda = +\infty :\; Q=2\pi
\end{equation}
makes the topological charge an angular variable.  Numerical work shows
that the potential energy is very nearly that of a pendulum, and that
$I=\xi M_s/m^2$ for some numerical constant $\xi$.  The resulting
approximate Hamiltonian has the form:
\begin{equation}     
H=M_s[\frac{\xi}{2M_W^2}\dot{Q}^2-\cos Q]
\end{equation}
which has, as it must, the value $M_s$ when $\dot{Q}=0,\;Q=\pi$.

Next one replaces $M_W$ by its time-dependent value, as in equation (7).
We have numerically investigated such driven pendulum equations.  They lead
to multiple transitions over the barrier, but we will not display such
solutions here.  One reason is that the {\it ansatz} we use here is
strictly tied to a unit change of topological charge, so that all that
counts is the rate of making a single transition over the barrier.  Just as
for all the classical barrier-hopping solutions presented for the $\phi$
{\it ansatz}, the rate is $O(\omega)$, very much different from the
tunneling rate.  (The tunneling rate is also changed as the sphaleron mass
oscillates; see \cite{gg}.)

To go further than this for a condensate of real sphalerons is
extraordinarily complicated; each sphaleron, like the instanton to which it
corresponds, has numerous degrees of freedom.  Even if we restrict this to
one degree of freedom (corresponding to $\lambda$) for each sphaleron, it
is not clear how to proceed.  Nor is it clear how to modify known
multi-instanton {\it ans\"atze} such as ADHM or that of 't Hooft or Jackiw,
Nohl, and Rebbi \cite{jnr} to express the real-time sphaleron dynamics in
the presence of an oscillating Higgs field.

\section{Conclusions}

In this work we have investigated two new mechanisms driven by preheating
oscillations of, {\it e.g.}, the Higgs field in hybrid inflation.  The
first mechanism, resonant barrier-crossing from a lower minimum to a higher
minimum (where there is no longer resonance), may explain some puzzles
associated with the symmetry-breaking patterns of GUTs.  This kind of
transition could also populate a metastable SU(3)$\times$SU(2)$\times$U(1)
vacuum in a supersymmetric extension of the Standard Model even if the
global minimum of the potential breaks charge and color. (In the case of
the MSSM, this posibility has direct inplications for collider
experiments~\cite{kls}.)  The second mechanism, resonant barrier-crossing
associated with B+L violation, may lead to a condensate of sphalerons on
time scales short compared to tunneling rates.  Both effects require
resonance with preheating oscillations to be effective.  We have not tried
to construct ``realistic" applications of these mechanisms to specific
preheating scenarios.  We note, however, that in many cosmological models,
even if the initial conditions are far from resonance, the system evolves
and reaches the resonance eventually, thanks to a change in the relevant
parameters~\cite{preheating}.  Such evolution is facilitated by either
non-quadratic inflaton potential that causes a variation in the inflaton
frquency, of by expansion of the universe and the associated Hubble damping
(for GUT, not weak-scale preheating), or some other effects that can slowly
drive a system into a resonance band.  We leave the building of realistic
cosmological models for future work.

Aside from such applications, there is still a good deal of work to be done
to clarify these mechanisms.  In the case of B+L violation, one can raise
the following issues:
\begin{enumerate}
\item How do the three stages (spatially-homogeneous potential, linear
momentum-mode perturbations, sphaleron condensate) of Section III evolve
from the first to the last?  This can only be answered by numerical work
more extensive than we have yet done.
\item The large-scale EW CS density we propose will have a projection onto
Maxwell magnetic fields carrying helicity (another term for Chern-Simons
number).  The spatially-homogeneous nature of these fields makes them quite
different from earlier proposals (see \cite{co97,fc} and references
therein) involving generation of Maxwell fields in a thermal environment,
with unacceptably small scale lengths to correspond to the scale lengths of
present-day galactic magnetic fields.  Given sufficient inverse cascading
of the nearly-homogeneous Maxwell fields following from our preheating
mechanism (at EW time these fields must be limited in extent by the Hubble
size), \cite{fc} shows that EW-time Maxwell fields could indeed be the
seeds for presently-observed galactic fields.  We intend to investigate
this further.
\item Can one make use of multi-instanton {\it ans\"atze} such as those of
ADHM, 't Hooft, or Jackiw, Nohl, and Rebbi \cite{jnr} to extend the
Bitar-Chang \cite{bc} construction we have exploited in Section IIIC in
order to understand quasi-analytically the formation of a sphaleron
condensate?
\item Are there (necessarily spin-dependent) quasi-resonant phenomena for
the production of W-bosons by an oscillating Higgs field which are in any
sense analogous to the very sharp resonant phenomena found by Cornwall and
Tiktopoulos \cite{ct} for spin-1/2 charged particles in specific
time-dependent electric fields?
\end{enumerate}

To clarify this last point, Ref. \cite{ct} found that it is possible to
have highly-resonant $e^+e^-$ pair production in a classical time-varying
electric field of the proper time dependence.  The sharply-resonant nature
of the process can only happen for fermions, but in any case spin effects,
which might be available with gauge bosons, are important in overcoming the
typical $\exp (-1/\alpha)$ rate of pair production in classical fields.

\acknowledgements The work of A. Kusenko was supported in part by
the U. S. Department of Energy under grant DE-FG03-91ER40662, Task C.

\appendix
\section*{Analysis of mode equations for $\phi$}

Here we give the analysis of the Mathieu-like but non-linear modal
equations of Section III.  Just as for the Mathieu equation, we write the
non-dimensionalized $\phi $ in the form
\begin{equation}   
\phi = a(t)\cos (rt/2)+b(t)\sin (rt/2),
\end{equation}
(where, as in the main text, $r=\omega /m$),  leaving out all terms with
higher frequencies.  One verifies that the time dependence of $a,b$ is
$O(\epsilon)$, so that we can ignore second derivatives of these
quantities.  However, we will save the cubic non-linearities.

Using equation (A1) in the equation of motion (8), saving only terms
varying as $\cos (rt/2)$ and $\sin (rt/2)$, and dropping second derivatives
yields:
\begin{equation}   
r\dot{a}+b[\frac{r^2}{4}+\frac{1}{2}\epsilon -1-\frac{3}{2}(a^2+b^2)]=0;
\end{equation}
\begin{equation}    
-r\dot{b}+a[\frac{r^2}{4}-\frac{1}{2}\epsilon -1-\frac{3}{2}(a^2+b^2)]=0.
\end{equation}
To make contact with the linear Mathieu equation, let us temporarily
replace the terms $(3/2)(a^2+b^2)$ by constants, and define an effective
(non-dimensional, that is, scaled by $m$) mass $M$ by:
\begin{equation}   
M^2\equiv 1+\frac{3}{2}(a^2+b^2).
\end{equation} 
Assuming exponential growth, with $a,b\sim\exp (\mu t)$, gives:
\begin{equation}    
mu=\frac{1}{2r}[\epsilon^2-(r^2-4M^2)^2]^{1/2}
\end{equation}
which gives growth only when $r=2M+O(\epsilon )$.  For small initial values
of $\phi$ this means $r\simeq 2$, but as $\phi$ grows because of the
initial parametric resonance, the system goes out of resonance.

We show that equations (A2,A3) can be solved exactly in terms of elliptic
integrals.  Multiply (A2) by $a$ and (A3) by $-b$ and add to get:
\begin{equation}    
\frac{d}{dt}(a^2+b^2)=-(\frac{2\epsilon}{r})ab.
\end{equation} 
This equation is independent of the non-linear terms in (A2,A3); it would
hold even if these terms were dropped.  Note that exponential growth
requires $a,b$ to be of opposite sign.  The constraints expressed by
equation (A6) allow the elimination of one degree of freedom:
\begin{equation}   
a=A\cos \Psi,\;\;b=-A\sin \Psi,
\end{equation}  
with a relation between $A$ and $\Psi$: 
\begin{equation}    
A=A_0\exp \int_0^t dt^{\prime}(\frac{\epsilon}{2r})\sin 2\Psi (t^{\prime})
\end{equation}
with $A_0$ as an initial value.  In the linear (Mathieu) case $\cos \Psi$
is the constant $|r^2-4|/\epsilon$, which yields the linear growth rate in
equation (A5).  But equations (A2,A3) yield two equations for the time
evolution of $A<\Psi$.  The sum of these equations is a trivial identity,
while the difference (using equations (A7,A8)) is:
\begin{equation}    
\epsilon \cos 2\Psi -2r\dot{\Psi}=\frac{r^2}{2}-2-3A_0^2\exp \int_0^t
dt^{\prime}(\frac{\epsilon}{2r})\sin 2\Psi (t^{\prime}).
\end{equation}
Now differentiate (A9) and use (A9) in the result to arrive at:
\begin{equation}   
2\ddot{\Psi}-\frac{\epsilon}{2r^2}(r^2-4)\sin
2\Psi+\frac{\epsilon^2}{2r^2}\sin 4\Psi .
\end{equation}
This is readily checked to be an elliptic equation.  We will not bother to
study it here.  All the physics is contained in the linearization of (A10),
which gives:
\begin{equation}    
\Psi(t)=\Psi_0\cos [(\lambda (t-t_0)]
\end{equation}
with frequency
\begin{equation}   
\lambda =[\frac{\epsilon^2}{r^2}-\frac{\epsilon (r^2-4)}{2r^2}]^{1/2}.
\end{equation}
Since $r^2-4$ is $O(\epsilon )$, so is $\lambda$.  
The best case for growth is 
\begin{equation}     
\Psi(t)=(\frac{\pi}{4})\cos \lambda t
\end{equation}
(so that $A$ and $b$ are equal initially).  Evidently, from (A8) growth
stops when $\Psi=0$, or when $t=\pi /2\lambda$.  This means, as discussed
in the main text, that growth cannot be unlimited.  (However, when the
frequency $r$ grows secularly, growth can continue unimpeded with
$a^2,b^2\sim r$, which maintains the resonant growth condition.)
Generally, no matter how small the initial values of the potential $\phi$,
eventually $\phi$ becomes of order unity.  The smaller the initial value,
the longer this process takes.


\end{document}